\documentstyle[prl,aps]{revtex}
\twocolumn
\begin{document}
\draft
\twocolumn[
\hsize\textwidth\columnwidth\hsize\csname@twocolumnfalse\endcsname
\title{Quasiparticle Localization in Disordered $d$-Wave Superconductors
}
\author{Jian-Xin Zhu,$^{1}$ D. N. Sheng,$^{1,2}$ and C. S. Ting$^{1}$
}
\address{ $^1$
Texas Center for Superconductivity and Department of Physics, 
University of Houston, Houston, Texas 77204\\
$^2$ Department of Physics and Astronomy, 
California State University Northridge, Northridge, CA 91330-8268 
}
\maketitle
\begin{abstract}
An extensive numerical study is reported on disorder effect in two-dimensional 
$d$-wave superconductors with random impurities in the unitary limit. It is 
found that a sharp resonant peak shows up in the density of states at zero 
energy and correspondingly the finite-size spin conductance is strongly 
enhanced which results in a non-universal feature in one-parameter scaling.  
However, all quasiparticle states remain localized, indicating that the 
resonant density peak alone is not sufficient to induce delocalization. 
In the weak disorder limit, the localization length is so long that  
the spin conductance at small sample size is 
close to the universal value predicted
by Lee  (Phys. Rev. Lett. {\bf 71}, 1887 (1993)).
\end{abstract} 
\pacs{PACS numbers: 74.25.Jb, 72.15.Rn, 74.62.Dh}
]

\narrowtext
Since the discovery of the $d$-wave pairing symmetry in high-$T_c$ 
cuprates, there has been increased interest in low-energy quasiparticle 
properties in unconventional superconductors. In such a $d$-wave 
superconductor, quasiparticles are gapless along the four nodal directions 
on the essentially cylindrical Fermi surface, in contrast to the 
conventional $s$-wave superconductors, where quasiparticles are gapped 
everywhere. The issues of how the disorder affects the low energy 
quasiparticle excitations and whether these quasiparticles are localized 
remain unresolved. Some perturbative self-consistent $T$-matrix (SCTM) 
calculations~\cite{GK85,SMV86,HVW86,Lee93,HL93,LW95,XW95,Joynt97} 
and a non-perturbative one~\cite{ZHH96} predicted a nonzero constant density 
of states (DOS) in low energy region in the presence of weak disorder. 
However, most of non-perturbative calculations~\cite{NT97,Mudry96} and one 
numerical study~\cite{AHM00} showed that the DOS at zero energy vanishes. 
With this constant or even vanishing DOS, some  
groups~\cite{Lee93,SFBN98} suggested that all quasiparticle states are
localized. On the other hand, it has been shown~\cite{BSR95} that a single
unitary 
scattering impurity produces a zero-energy quasiparticle resonant state 
while the long-range overlap between these impurity states~\cite{BS96} may 
lead to extended quasiparticle band near zero energy. More recently, a 
singularity in the DOS at zero energy was obtained by non-perturbative 
$T$-matrix method~\cite{PL00} for the random distributed unitary impurities.
It is noteworthy that in one dimension, there is a direct relation between 
the localization length and the DOS~\cite{Thouless72}; thus a singularity 
in zero-energy DOS signals the delocalization in the system. However, 
in two dimensions, this theorem does not hold generally~\cite{PL00}. For 
example, in two dimensional integer quantum Hall systems, it has been shown 
that the delocalization property at the quantum critical point is not changed 
by the changing of the DOS due to strong electron-electron 
interaction~\cite{YMH95}. Therefore, the localization of quasiparticles in a 
$d$-wave superconductor in the presence of nonmagnetic unitary impurities is 
still an open question.  

In this Letter, we numerically examine the disorder effect in $d$-wave 
superconductors with nonmagnetic impurities in the unitary limit. 
The quasiparticle DOS is calculated by exact diagonalization and the
spin conductance is computed by the transfer matrix method. It is found 
that, depending on the particle-hole symmetry of the Hamiltonian, a sharp 
DOS peak can occur at zero energy and correspondingly the spin conductance 
is strongly enhanced at finite sample size. However, using one parameter 
scaling analysis we show that all the quasiparticle states are always 
localized regardless of the existence of the zero-energy peak in the DOS. 
In weak disorder limit, the localization length is so long that the spin
conductance at small sample size remains 
close to the universal value $2\xi_0/a $ ($\xi_0$ is the coherence length 
of the superconductor and $a$ is the lattice constant) in agreement with
the theoretical prediction by Lee~\cite{Lee93}. A non-universal feature
in one-parameter scaling of conductance related to the resonant DOS peak
is also discussed.

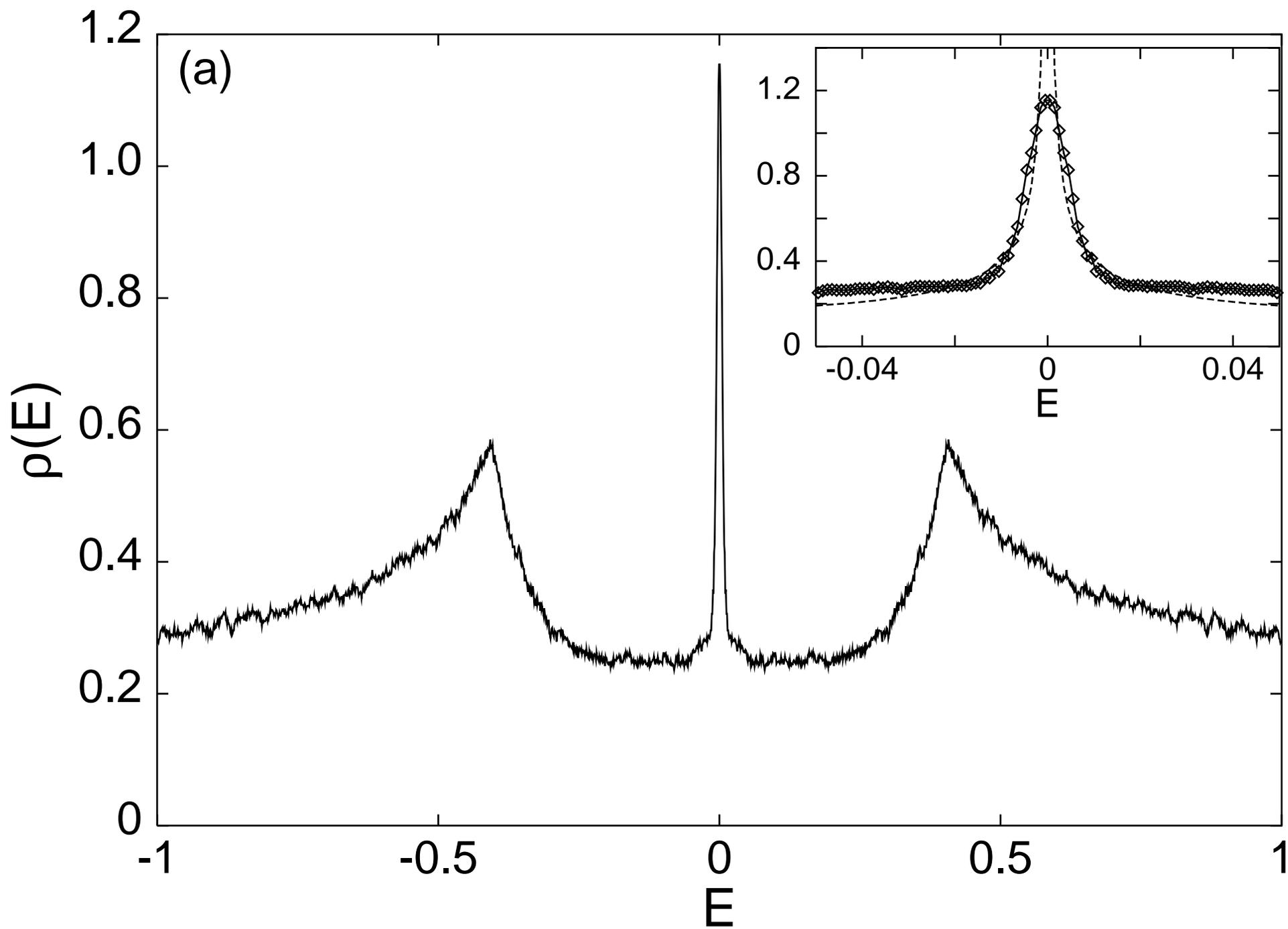
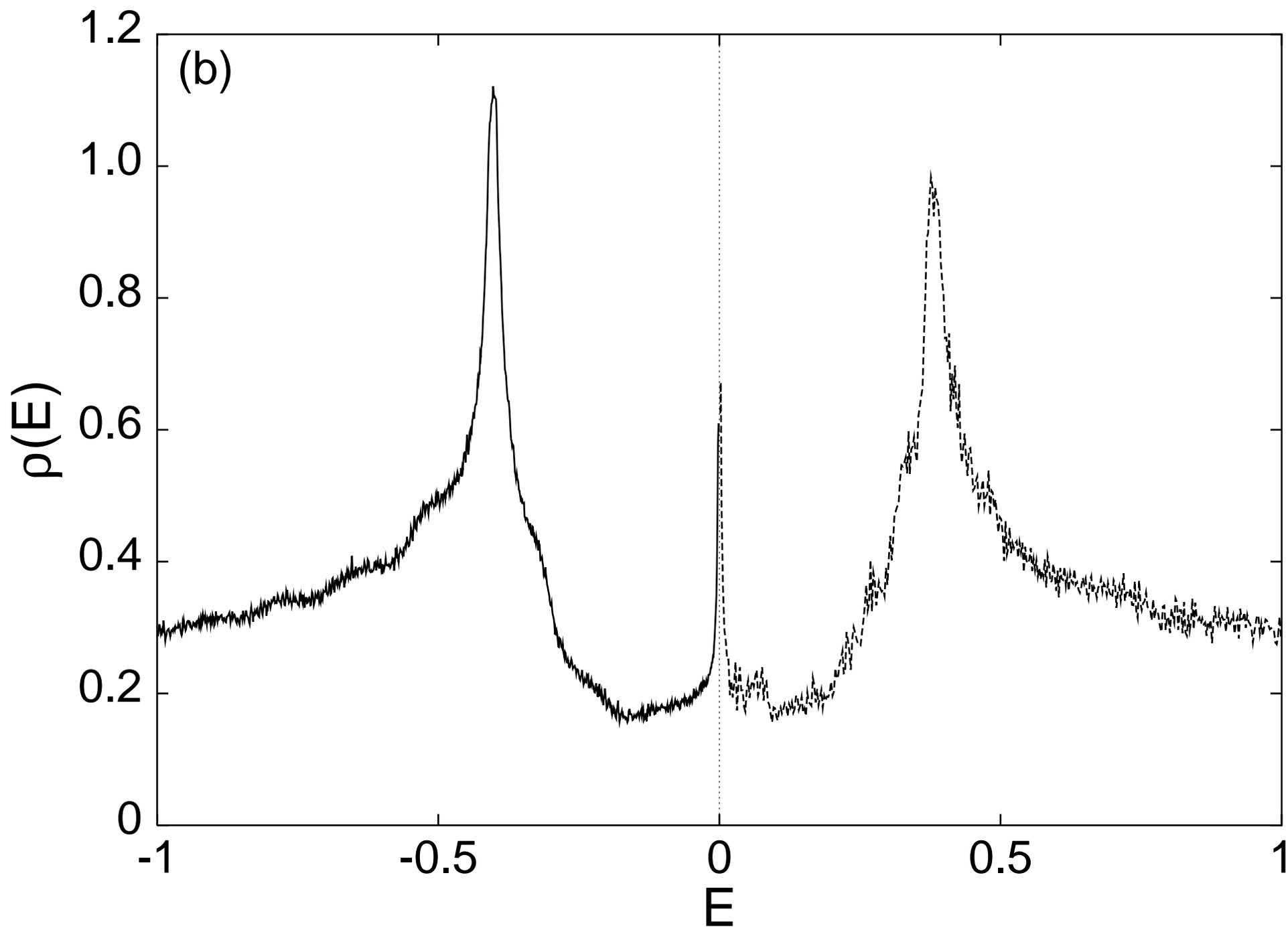
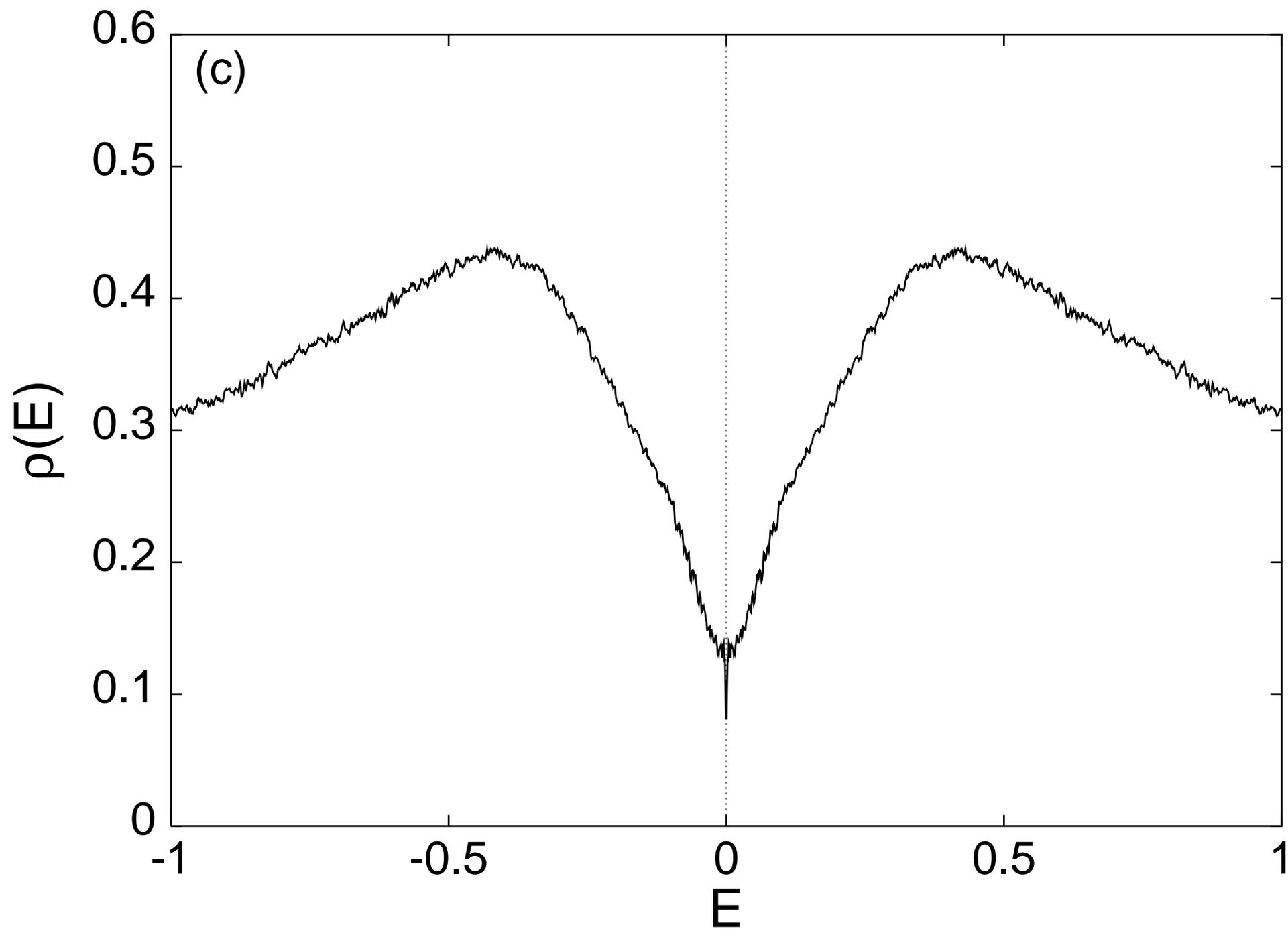
\begin{figure}
\caption[long]{(a) Density of states $\rho(E)$ in a $d$-wave 
superconductor (sample size $90\times 90$)
with a fraction $n_i=0.1$ of the randomly distributed 
impurities in the unitary scattering limit $U_0=100$. 
Inset of (a): The DOS peak is fitted using the analytical form
$c_{0}n_{i}/2|E|(\ln^{2}|E/\Delta_g|+\pi^2/4)$ ($E_g=4\times \Delta_d$) 
from Ref.~\cite{PL00} with $c_0=0.66$.  
(b) Comparison of $\rho(E)$ ($E<0$ part) at  $N_L=90 \times 90$ 
for $n_{i}=0.04$   with the self-consistent result ($E>0$ part) at
$N_L=25\times 25$ (averaged over $8\times 8$ supercells).
(c) $\rho(E)$ at $N_L=90\times 90$ with on-site random disorders
distributed between $[-1,1]$.   
} 
\label{FIG:DOS} 
\end{figure}

We begin with a lattice Hamiltonian for the $d$-wave 
superconductor~\cite{FB99}
\begin{eqnarray}
H&=&-\sum_{\langle ij\rangle,\sigma  } 
c_{i\sigma}^{\dagger}c_{j\sigma}
+\sum _{i,\sigma} (U_i-\mu) c^{\dagger}_{i\sigma} 
c_{i\sigma} \nonumber \\ 
&&+\sum_{\langle ij\rangle}   
[\Delta _{ij} c_{i\uparrow}^{\dagger}c_{j\downarrow}^{\dagger} 
+ \mbox{h.c.}]\;,
\label{EQ:Hamil}
\end{eqnarray}
where $\langle ij\rangle $  refers to two nearest neighboring sites 
with  the hopping integral taken as the unit,  $\mu$ is the chemical potential 
and $U_{i}$ is the impurity potential. We mainly consider the unitary limit 
where $U_i$ takes a nonzero value $U_0$ only at a fraction $n_i$ of the sites 
which are randomly distributed in space.  The $d$-wave symmetry is imposed 
by choosing order parameters: 
$\Delta_{i,i\pm \hat {x}}=-\Delta_{i,i\pm \hat{y}}=\Delta_{d}$, which 
yields the excitation spectrum $E_{k}=\sqrt{\epsilon_{k}^{2}+\Delta_{k}^{2}}$, 
with $\epsilon_{k}=-2(\cos k_{x} +\cos k_{y})-\mu$ and
$\Delta_{k}=2\Delta_{d}
(\cos k_{x}-\cos k_{y})$. 
Therefore, gapless quasiparticle states exist along the direction $\vert 
k_{x}\vert =\vert k_{y}\vert$ in the momentum space. Unless otherwise stated, 
$\mu=0$, $\Delta_d=0.1$, and $U_0=100$ are taken throughout the work. In the 
presence of a single impurity in the unitary scattering limit, earlier study 
has shown~\cite{ZLTH00} that the order parameter is strongly suppressed 
near the impurity site on a scale of a few lattice constant. To take into 
account this effect, the order parameters on bonds connecting with the strong 
impurity sites are taken as zero. By exactly diagonalizing the Hamiltonian (1), 
one can calculate the quasiparticle DOS, which is defined as~\cite{AHM00,note1} 
\begin{equation}
\rho(E)=\frac{1}{N_L}\sum_{n}\delta(E-E_{n})\;,
\end{equation}
where $N_L=L\times L$ with $L$ the linear dimension of the system in units 
of lattice constant ($a=1$).  In Fig.~\ref{FIG:DOS}(a), the DOS is plotted  as a function of quasiparticle 
energy $E$ with impurity density $n_i=0.1$ ($\mu=0$ and $\Delta_{d}=0.1$) 
at  $N_L=90\times 90$. In the calculation the periodic boundary condition 
is used and it has been checked that the results do not depend on the boundary 
condition for large $L$ considered here. As shown in Fig.~\ref{FIG:DOS}(a), 
we find that a sharp zero-energy peak shows up in the DOS, which can be fitted 
by the analytical form $c_0 n_i/2|E|(\ln^{2}|E/E_g|+\pi^2 /4)$ 
from Ref.~\cite{PL00} with $c_0=0.66$ (the superconducting gap $E_g \sim 4
\times \Delta _d$) as shown in the inset of Fig.~\ref{FIG:DOS}(a).  The 
fitting breaks down at an  energy scale close to $1/N_L$~\cite{note2}.
The strength of the zero-energy peak is reduced when $n_i$ is changed 
to $0.04$ as shown in $E<0$ part of Fig.~\ref{FIG:DOS}(b).  The overall 
shape of the peak is sample size independent (from $N_{L}=25\times 25$ to 
$120\times 120$ as well as a strip system $30\times 300$) with the DOS 
value at the peak position increasing 
with $N_L$ very slowly. Note that in the presence of disorder, the order 
parameter $\Delta _{ij}$  is in principle subject to the self-consistency 
condition: $\Delta_{ij}=-g_{ij}\langle c_{j \downarrow}c_{i\uparrow}\rangle $
where $g_{ij}=g_0$ is the attracting interaction for $d$-wave
pairing. We have also calculated the DOS by diagonalizing the 
Hamiltonian Eq.~(\ref{EQ:Hamil}) self-consistently 
for each disorder configuration with  $N_{L}=25 \times 25$. The obtained DOS 
is also shown in Fig.1(b) ($E>0$ part), which is averaged over 50 impurity 
configurations and $8\times 8$ wavevectors in the supercell Brillouin zone. 
As can be seen,  the DOS value at $E_g$ is reduced due to the suppression of 
$\Delta_{ij}$ around each impurity site. However, all other features remain 
essentially unchanged (compared to the $E<0$ part of Fig. 1(b)). 

The presence of zero-energy peak in the DOS crucially depends on the 
symmetry of the Hamiltonian. Since a repulsive or attractive impurity 
center with infinite strength (i.e., unitary limit) is equivalent to the 
exclusion of a lattice site, both the 
local and global particle-hole symmetry remain
if the chemical potential $\mu=0$, which then produces a resonant peak at 
$E=0$~\cite{PL00}. This feature was not exhibited in earlier works 
because of either the breaking of local particle-hole symmetry by the soft 
impurity scattering~\cite{NT97,Mudry96,SFBN98} or the breaking of the band 
particle-hole symmetry by considering $\mu\neq 0$~\cite{AHM00}, which 
indicates the importance of the realization of disorder model.  
In Fig.~\ref{FIG:DOS}(c), the DOS is presented for 
random on-site disorders with $U_i$ uniformly  distributed between 
[-1,1] with a homogeneous order parameter at all sites. 
Due to the absence of the local  particle-hole symmetry in this case, 
$\rho(E)$ has a finite value at low energy region down to a mesoscopic 
scale $E\sim 1/N_L$, below which it shows a zero-energy dip, in agreement 
with those of the non-perturbative calculations~\cite{NT97,Mudry96,SFBN98}.  
We have also relaxed the unitary scattering limit by 
taking $U_{0}=10$ (comparable to the band width) or (and) 
broken the band particle-hole symmetry by taking $\mu=-1$, 
and found that the DOS at $E=0$ is always strongly suppressed, 
which is similar to the results obtained in Ref.~\cite{AHM00}. 
There is a smooth crossover from the zero-energy peak to dip in the 
DOS with the varying of model parameters as long as the Hamiltonian is 
driven away from the perfect particle-hole symmetry. 
Therefore, our numerical result indicates that different realizations of
disorder give rise to different profiles of the DOS  as the energy
approaches to the Fermi level.
 
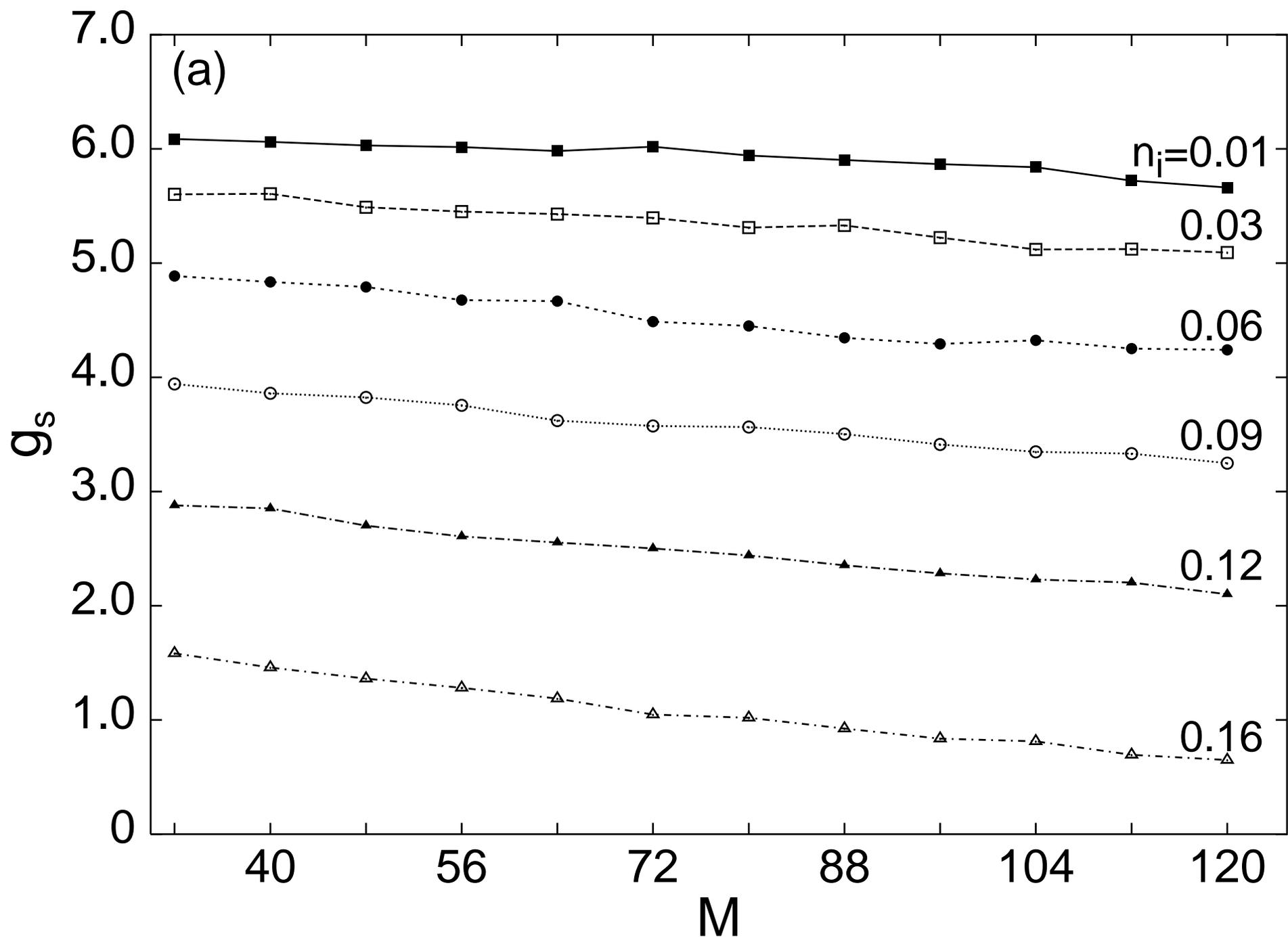
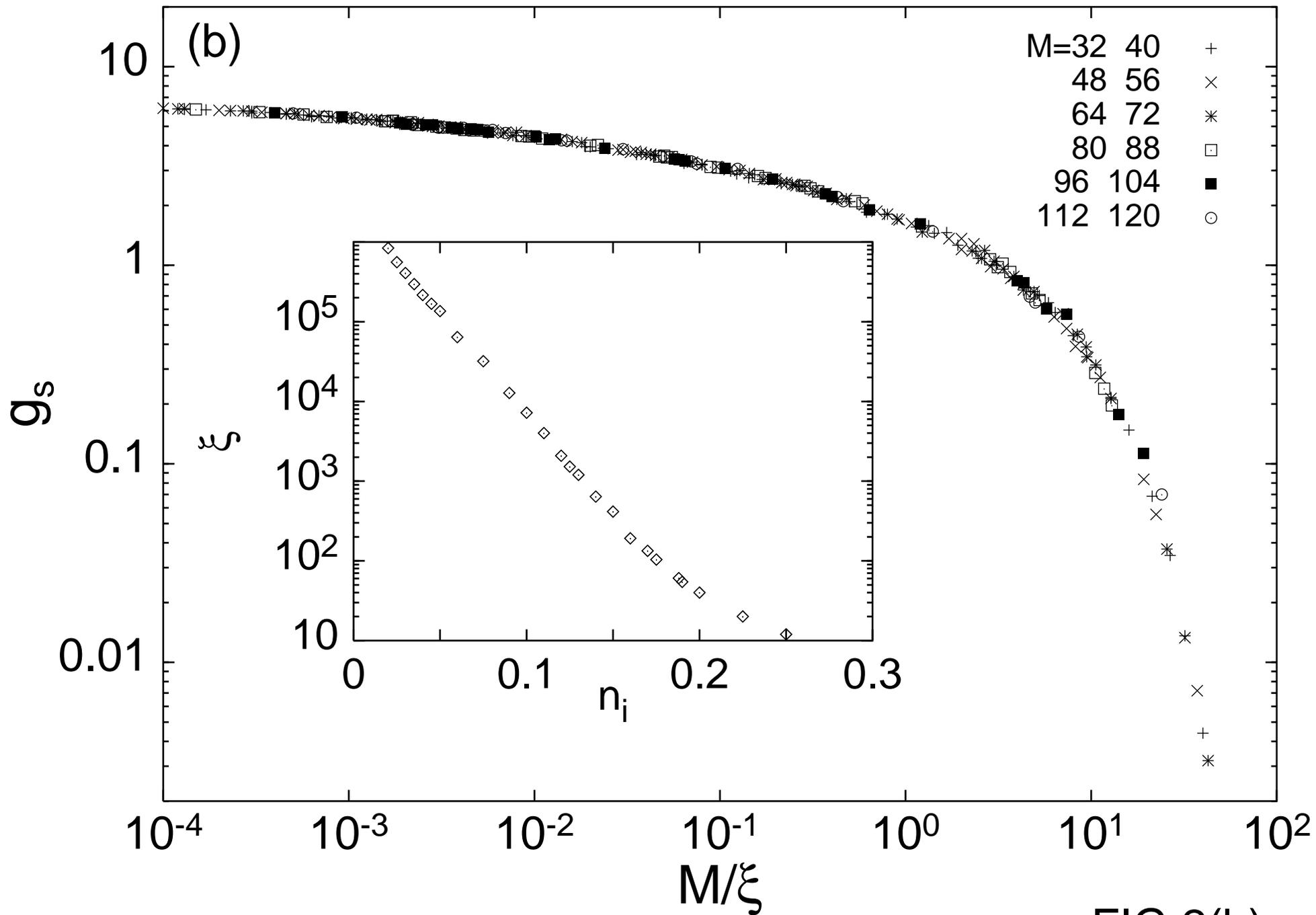
\begin{figure}
\caption{
(a) Spin conductance $g_{s}$ as a function of the strip width $M$ for 
different impurity density $n_i$ at $E=0$. (b) Double logarithmic 
plot of  $g_s$ as a scaling function of $M/\xi$ at $E=0$ for all
the data with $0.01\leq n_{i}\leq 0.275$.
Inset of (b): The scaling parameter $\xi$ as a function of  $n_i$ at $E=0$. 
The other parameters are  $\mu=0$, $\Delta_d=0.1$, and $U_0=100$.
} 
\label{FIG:SCALING}
\end{figure}

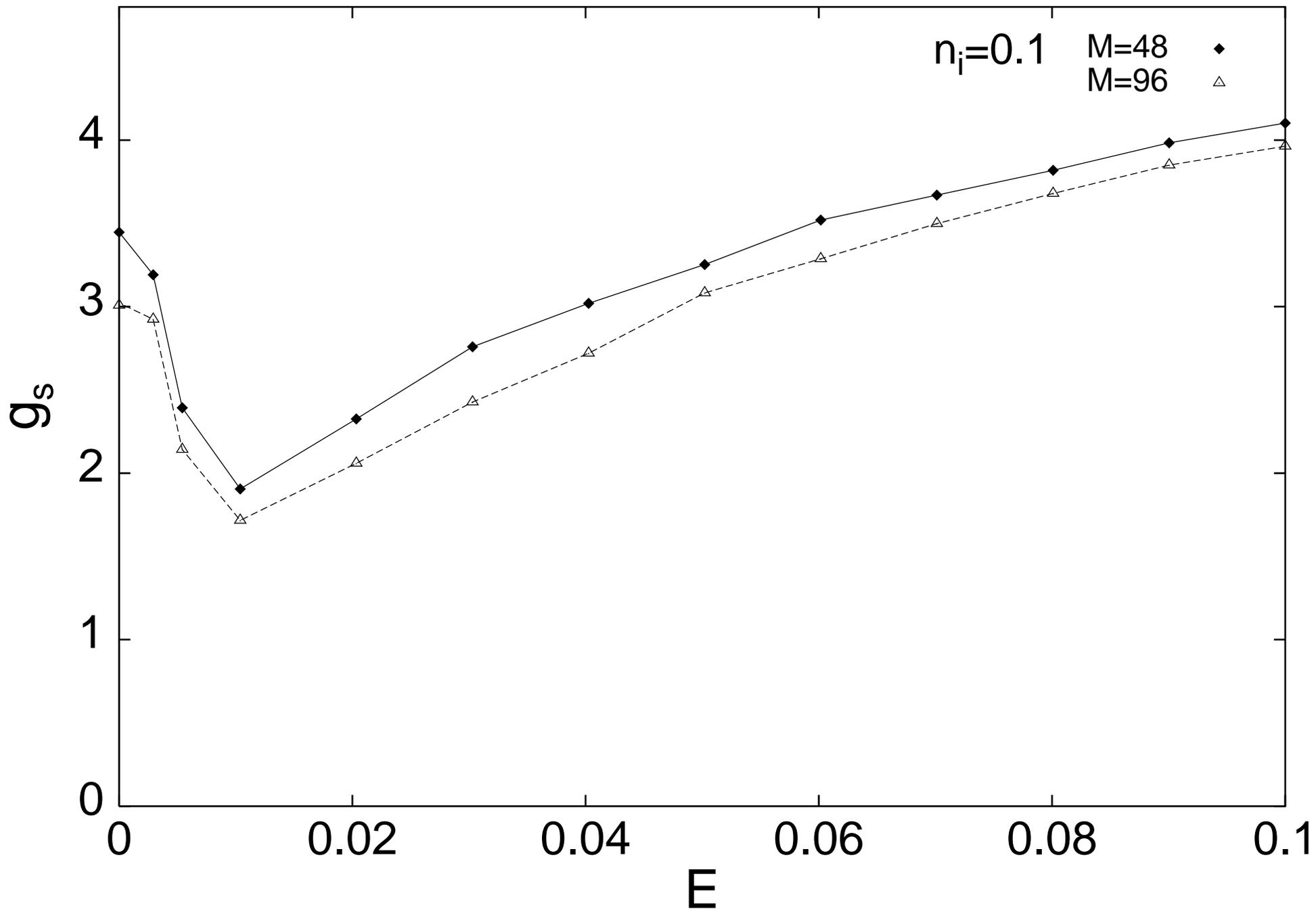
\begin{figure}
\caption{$g_s$ as a function of quasiparticle energy $E$ at $M=48$ and $96$ 
for $n_i=0.1$.  All the other parameters are the same as in Fig.2.  
} \label{FIG:SPC}
\end{figure}

In the absence of the zero-energy resonant peak in DOS 
due to the breaking of band particle-hole symmetry by $\mu\neq 0$,
Franz {\em et al.}~\cite{FKB96} have studied the similar problem 
by examining the sensitivity of the wave function to the boundary 
conditions and by analyzing the finite-size dependence of inverse 
participation ratios, and presented a strong evidence for the 
localization of low energy quasiparticles. 
Here we are concerned with the question of quasiparticle 
localization or delocalization in the disordered $d$-wave superconductor 
when the zero-energy peak appears in the DOS. 
We employ the transfer-matrix method 
to calculate the finite-size localization length and  the longitudinal 
conductance. We consider the quasi-one-dimensional strip sample with  
the length $L\ge 10^5$ and width $M$. The quasiparticle wavefunction 
amplitudes in the $ix$-th and $(ix+1)$-th slices satisfy the following 
equation: 
\begin{equation}
\left( \begin{array}{c}
\hat{\phi}_{ix+1} \\
\hat{\phi}_{ix} 
\end{array}
\right) = T_{ix} 
\left( \begin{array}{c}
\hat{\phi}_{ix} \\
\hat{\phi}_{ix-1} 
\end{array}
\right)
\;,
\end{equation}
where $\hat{\phi}_{ix}$ is a $2M$-component vector of the 
Bogoliubov amplitudes for quasiparticle states, and $T_{ix}$ is a 
$4M\times 4M$ transfer matrix. The transfer matrix through the whole system, 
$P_{L}=\prod_{ix=1}^{L} T_{ix}$, has a set of $2M$ pairs of Lyapunov 
exponents, which determine the inverse of length $\lambda_{i}$  
($i=1,2,\dots,2M$). The usual  orthonormalization procedure is 
taken~\cite{MK83} in our calculation.  Correspondingly, the longitudinal 
conductance $g_s$ extrapolated for square sample with width $M$ is given 
by~\cite{Beenakker97}:
\begin{equation}
g_{s}(M,n_{i})=\sum_{j=1}^{2M} \cosh^{-2}\Lambda_{j}\;,
\end{equation}
where $\Lambda_{j}=\lambda_{j}/M$.  Note that $g_s$ corresponds to the 
spin conductance as the spin carried by quasiparticle is 
conserved~\cite{SFBN98}. As shown in Fig.~\ref{FIG:SCALING}(a), $g_s$ as a 
function of $M$ monotonically decreases with the increase of $M$ at $E=0$ for 
each selected impurity density from $n_i=0.01$ to $0.16$,  consistent with
localization in the large $M$ limit.  All the data between $M=32$ and $M=120$ 
at different $n_i$  can be collapsed onto a single curve: 
\begin{equation}
g_{s}(M,n_i)=f\left(\frac{\xi(n_i)}{M}\right)\;,
\end{equation}
as shown in Fig.~\ref{FIG:SCALING}(b) in accordance with the one-parameter
scaling law.  Here $\xi(n_i)$ is the thermodynamic localization 
length which only depends on $n_i$  as shown in the inset of 
Fig.~\ref{FIG:SCALING}(b), and it remains finite for all the disorder 
density $n_i$, suggesting that all the states are localized even in the 
unitary limit with the presence of the zero energy resonant peak. 
In addition, we display in Fig.~\ref{FIG:SPC} the conductance $g_s$  
as a function of quasiparticle energy $E$ at sample width $M=48$ and $96$ 
with $n_i=0.1$. It is found that $g_s$ is strongly enhanced as  
$E\rightarrow 0$ in the region $E\sim 0.01$ corresponding to the width of 
the DOS peak while away from this region $g_s$ generally increases with the 
increase of $E$.  For quasiparticle at low energy with $E<0.1 $, it has been 
found that $g_s$ always decreases with the increase of $M$ and all the states 
are localized in large $M$ limit.  However, the scaling curve 
$g_s(M,n_i)=f(\frac{\xi(n_i)}{M})$ found at $E$ away from the DOS peak 
($E>0.01$) is  different from that for $E=0$ at the peak of the DOS, 
indicating the breaking-down of the universal one-parameter scaling 
law due to the presence of the resonant peak in the DOS. 
In addition, for all values of $\Delta_d$, $g_s$ 
decreases with $M$, implying localization in all the parameter region.  
At fixed $M$, $g_s$ always decreases monotonically with the increasing 
$\Delta_d$, in agreement with the  
general argument that quasiparticle states in a superconducting phase
are always more localized~\cite{Lee93} than the corresponding normal state.  
Given the localization of the quasiparticle states, the spin conductance
at the longest length scales must vanish. 
However, in the weak disorder (the impurity density $n_i \le 4\%$), i.e.,
the Born limit, we found that the localization length is so long that at
small sample  size $M\sim 32$, the spin conductance  $g_s^0$ follows the
universal form~\cite{Lee93} $g_s^0=2\xi_0/a$  as the coherence length
$\xi_0 $ is changed from $1.6 $ to $6.4$ by changing $\Delta_d$ 
between $0.05$ and $0.20$.
For the strong disorder, the conductance is generally
smaller than the Born limit value due to the onset of localization effect.

In conclusion, we have studied the quasiparticle states in 2D $d$-wave 
superconductors with randomly distributed strong impurities in the unitary 
scattering limit. As the particle-hole symmetry holds, a very sharp DOS peak 
is obtained at zero energy. Such a DOS peak enhances the finite-size 
conductance. However, using one-parameter scaling analysis  
we have shown that all the quasiparticle states at low energy are still
localized 
and the localization effect is generally enhanced with the increase
of superconducting order parameter.

{\bf Note added}: After the submission of this paper, we received 
a preprint from Atkinson {\em et al.}~\cite{AHMZ00} where similar results 
for the DOS in the unitary limit of the symmetric band  were obtained.

{\bf Acknowledgments} - The authors would like to acknowledge 
A. V. Balatsky, M. P. A. Fisher,  P. A. Lee,  X.-G. Wen, and Z. Y. Weng for 
helpful and stimulating discussions. JXZ also thanks W. A. Atkinson for 
patiently explaining their work to him. 
This work is supported by the State of Texas through ARP Grant No. 3652707, 
the Texas Center for Superconductivity at 
University of Houston, and the Robert A. Welch Foundation.

\begin{references}
\bibitem{note1} Notice that the band DOS defined here is related to  
the local density of states as, $\rho(E)=\frac{1}{2N_{L}}\sum_{\bf i}
[{\cal N}_{\bf i}(E)
+{\cal N}_{\bf i}(-E)]$, where ${\cal N}_{\bf i}=2\sum_{n}(\vert 
u_{\bf i}^{n}\vert^{2} \delta(E-E_{n})
+\vert v_{\bf i}^{n}\vert^{2} \delta(E+E_{n})$
with $(u_{\bf i}^{n},v_{\bf i}^{n})$ the Bogoliubov 
amplitudes corresponding to the $n$-th quasiparticle eigenstate.
\bibitem{note2} Due to the finite-size  level repulsion effect, the 
DOS vanishes exactly at $E=0$. However, the integrated DOS at low energy 
region $|E|\sim 0.0005$ saturates to a constant in this case, suggesting the 
unimportance of the level-repulsion gap in the large $L$ limit.
\end{references}
\end{document}